\documentclass[prc,aps,twocolumn,floats,floatfix,showpacs,superscriptaddress,nofootinbib]{revtex4}
\usepackage{amsmath}
\usepackage{amsfonts}
\usepackage{graphicx}

\begin{document}

\title{Microscopic description of the pygmy dipole resonance
in neutron-rich Ca isotopes}

\author{N. N. Arsenyev}
\affiliation{Bogoliubov Laboratory of Theoretical Physics,
             Joint Institute for Nuclear Research,
             141980 Dubna, Moscow region, Russia}
\author{A. P. Severyukhin}
\affiliation{Bogoliubov Laboratory of Theoretical Physics,
             Joint Institute for Nuclear Research,
             141980 Dubna, Moscow region, Russia}
\affiliation{Dubna State University,
             141982 Dubna, Moscow region, Russia}
\author{V. V. Voronov}
\affiliation{Bogoliubov Laboratory of Theoretical Physics,
             Joint Institute for Nuclear Research,
             141980 Dubna, Moscow region, Russia}
\author{Nguyen Van Giai}
\affiliation{Institut de Physique Nucl\'eaire, CNRS-IN2P3 and
             Univ. Paris-Sud, 91405 Orsay, France}

\begin{abstract}
We study the effects of the phonon-phonon coupling on the
low-energy electric dipole response within a microscopic model based
on an effective Skyrme interaction. The finite rank separable
approach for the quasiparticle random phase approximation is used.
Choosing as an example the isotopic chain of Calcium, we show
the ability of the method to describe the
low-energy $E1$~strength distribution. With one and the same set of
parameters we describe available experimental data for~$^{48}$Ca and
predict the electric dipole strength function for $^{50}$Ca.
\end{abstract}


\maketitle

\section{Introduction}

With the fast development of radioactive beam techniques over
the past several decades, experiments focused on the exotic
short-lived nuclei were performed extensively. New interesting phenomena
arise in nuclei with a strong imbalance of the proton~(p) and neutron~(n)
numbers~\cite{Savran13}. One of the phenomena sensitive to the change
in $N\!/Z$~ratios is the pygmy
dipole resonance~(PDR). The PDR leads to the enhancement of dipole
strength well below the region of the giant dipole
resonance~(GDR). In analogy to the GDR, the PDR has
been interpreted as a collective
oscillation of the neutron skin, $\Delta R_{\rm np}$, with respect to a $N{\approx}Z$
inert core; see, e.g., Refs.~\cite{Paar07,RocaMaza18}. The
total sum of the measured energy-weighted sum rule~(EWSR) of this $E1$
distributions is less than $1{\div}2{\%}$ of the Thomas-Reiche-Kuhn~(TRK)
sum rule value for stable nuclei and less than $5{\div}6{\%}$ for
unstable neutron-rich nuclei~\cite{Savran13}. Nevertheless, currently
the structure and dynamics of the PDR is one of
the actively studied topics in nuclear
physics. One of the reason is that the PDR study is expected to
provide information on the symmetry energy term of the nuclear
equation of state~\cite{RocaMaza18}.

The study of new exotic nuclei (or/and modes of excitation)
stimulates the development of nuclear models to describe
properties of nuclei away from the stability valley. The
quasiparticle random phase approximation~(QRPA) with a
self-consistent mean-field derived from Skyrme energy density
functionals~(EDF) is one of the most successful methods for
studying the low-energy dipole strength, see e.g.,
Refs.~\cite{Paar07,RocaMaza18}. A description of the
low-energy $E1$~strength distribution requires to include
the coupling between one- and two-phonon
components of the wave functions~\cite{Soloviev78,Grinberg94}. The main difficulty is
that the complexity of calculations beyond
standard QRPA increases rapidly with the size of the configuration
space, so that one has to work within limited spaces.
With the finite rank separable approximation
(FRSA)~\cite{Giai98,Severyukhin08} for the residual interaction,
one can perform Skyrme-QRPA calculations in very large
two-quasiparticle~(2QP) spaces. Following the basic ideas of
the quasiparticle-phonon model (QPM)~\cite{Soloviev92},
the FRSA has been generalized for the
phonon-phonon coupling (PPC)~\cite{Severyukhin04}.

The FRSA was used while studing the electric
low-energy excitations and giant resonances within and beyond the
Skyrme-QRPA approach~\cite{Severyukhin04,Severyukhin12,Severyukhin17}.
In this paper, we discuss the PPC effect on the properties
of PDR in the Ca isotopes. We illustrate our approach with the
stable isotope~$^{48}$Ca having the closed neutron shell $N{=}28$,
in comparison with the unstable isotope~$^{50}$Ca with $N{=}30$. These
nuclides from the Ca chain are suitable candidates to follow
the PDR evolution. Our results for neutron-rich Ca and Sn isotopes were reported in
Refs.~\cite{Arsenyev12,Arsenyev15,Arsenyev17}.

\section{Brief outline of the FRSA model}

\begin{figure}[t!]
\includegraphics[width=\columnwidth]{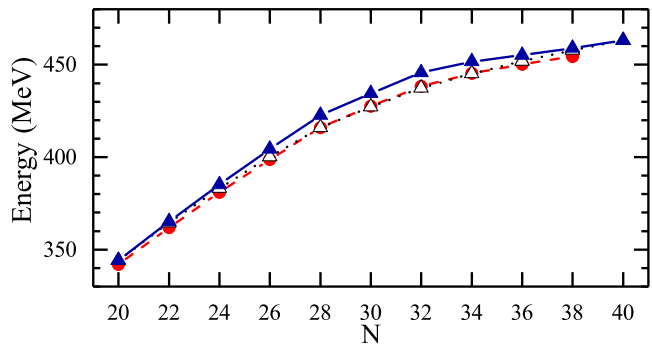}
\caption{(Color online) Binding energies of the even-even Ca
isotopes as a function of neutron number. Results of calculations
within the HF-BCS with the SLy5 EDF --- open triangles, with SLy5{+}T EDF ---
filled triangles. Experimental and extrapolated energies  from the AME2016 atomic
mass evaluation~\cite{Wang17} --- filled circles.}
\label{CaBE}
\end{figure}

The FRSA approach has been discussed in detail in
Refs.~\cite{Giai98,Severyukhin08,Arsenyev17} and we briefly present it
here  for completeness. The SLy5~\cite{Chabanat98} and
SLy5{+}T~\cite{Colo07} EDF are used in the
Hartree--Fock-BCS~(HF-BCS) calculations as well as for the
particle-hole~(\rm{p-h}) channel. The parameters of the Skyrme force SLy5
have been adjusted to reproduce nuclear matter properties, as well
as nuclear charge radii and binding energies of doubly magic nuclei.
The force SLy5{+}T involves tensor terms added without refitting
the parameters of the central interaction (the tensor interaction
parameters are $\alpha_{T}{=}-170$~MeV{}fm$^5$ and
$\beta_{T}{=}100$~MeV{}fm$^5$). These parameterizations enable to correctly describe
binding energies of even-even Ca isotopes. This is
illustrated in Fig.~\ref{CaBE}, where the calculated binding
energies for $^{40-60}$Ca together with experimental and extrapolated
data (AME2016)~\cite{Wang17} are shown. The agreement between the HF-BCS
results and data is reasonable, the deviations being less than
2{\%}.
For the interaction in the particle-particle~(\rm{p-p}) channel,
we use a zero-range volume force. The pairing strength
is taken equal to $-270$~MeV{}fm$^{3}$.
This value is fitted to reproduce
the experimental neutron pairing gaps of~$^{50,52,54}$Ca
obtained by the three-point formula~\cite{Severyukhin08,Arsenyev17}.
This kind of pairing interaction has allowed describe satisfactorily
experimental data for $^{70,72,74,76}$Ni~\cite{Severyukhin14}, $^{90,92}$Zr and
$^{92,94}$Mo~\cite{Severyukhin12}. Thus, hereafter we use the
Skyrme interaction SLy5 with and without tensor components in the
particle-hole channel together with the volume zero-range force
acting in the \rm{p-p} channel.

The residual interaction in the \rm{p-h} channel $V^{({\rm p-h})}_{\rm
res}$ and in the \rm{p-p} channel $V^{({\rm p-p})}_{\rm res}$ can be
obtained as the second derivative of the energy density functional
with respect to the particle density and the pair density,
accordingly. Following Ref.~\cite{Giai98} we simplify
$V^{({\rm p-h})}_{\rm res}$ by approximating it by the Landau--Migdal form.
Moreover, we neglect the $l{=}1$ Landau parameters (in the case of
the Skyrme EDFs, the Landau parameters with $l{>}1$ are equal to
zero). The Landau parameters $F_0$, $G_0$, $F^{\prime}_0$,
$G^{\prime}_0$ expressed in terms of the Skyrme force parameters
depend on the Fermi momentum $k_{\rm F}$ of nuclear
matter~\cite{Giai81}.

To take into account the effects of the PPC we follow the basic QPM
ideas~\cite{Soloviev92}. We construct the wave
functions of excited states as a linear combination
of one- and two-phonon configurations
\begin{gather}
 \begin{matrix}
 \Psi_{\nu}(\lambda\mu)=\Biggl(\sum\limits_{i}R_{i}(\lambda\nu)Q_{\lambda\mu i}^{+}\\
 +\sum\limits_{\lambda_{1}i_{1}\lambda_{2}i_{2}}P_{\lambda_{2}i_{2}}^{\lambda_{1}i_{1}}
 (\lambda\nu)\left[Q_{\lambda_{1}\mu_{1}i_{1}}^{+}Q_{\lambda_{2}\mu_{2}i_{2}}^{+}\right]_{\lambda\mu}\Biggr)|0\rangle,
 \end{matrix}
 \label{wf2ph}
\end{gather}
where $|0\rangle$~is the phonon vacuum, $Q_{\lambda\mu i}^{+}$ is
the phonon creation operator and $\nu$ labels the excited states.
The coefficients $R_{i}(\lambda\nu)$,
$P_{\lambda_{2}i_{2}}^{\lambda_{1}i_{1}}(\lambda\nu)$ and energies of the
excited states $E_{\nu}$ are determined from the variational
principle which leads to a set of linear
equations~\cite{Severyukhin04,Severyukhin12}. The equations have the same form
as in the QPM~\cite{Soloviev78,Grinberg94,Soloviev92}, but the
single-particle spectrum and the parameters of the residual
interaction are obtained from the chosen Skyrme EDFs without any
further adjustments. In order to let the two-phonon
components of the wave functions obey the
Pauli principle the exact commutation relations
between the phonon operators should be taken into account.
In the present case, i.e. for the dipole states which are constructed from
${\rm 1p{-}h}$~components corresponding to transitions between neighboring main shells
the Pauli principle corrections to coupling matrix elements with two-phonon configurations
consisting of low-lying phonons of different
multipolarities are small (see Ref.~\cite{Soloviev92}).

\begin{figure}[t!]
\includegraphics[width=\columnwidth]{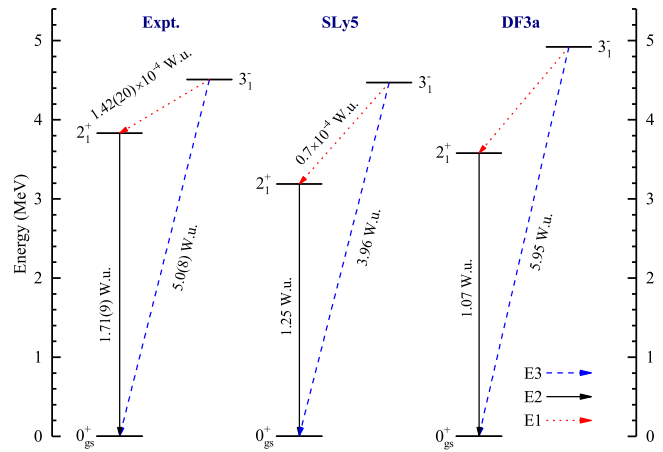}
\caption{(Color online) Experimental \cite{Burrows06} and
theoretical energies and transition probabilities of the
$[2_{1}^{+}]_{RPA}$ and $[3_{1}^{-}]_{RPA}$ states for $^{48}$Ca.
Results for the Skyrme EDF (SLy5) --- the present paper, those for the Fayans EDF
(DF3a) are taken from~\cite{Saperstein16}. $B(E1;\downarrow)$,
$B(E2;\downarrow)$, and $B(E3;\downarrow)$ factors
are given in Weisskopf units (W.u.).}
\label{Fig_Ca48sp}
\end{figure}
\begin{figure}[t!]
\includegraphics[width=\columnwidth]{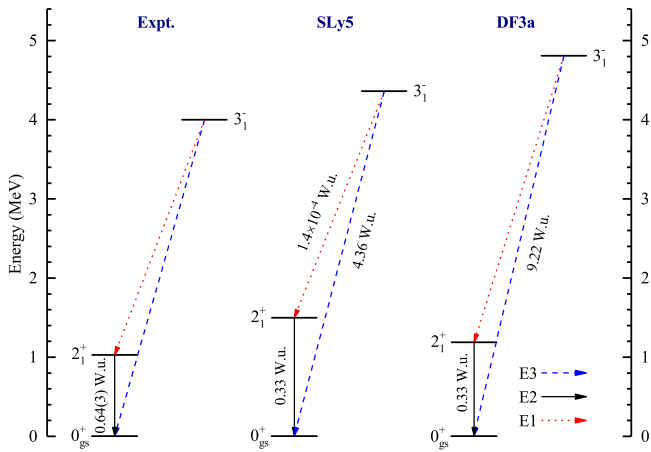}
\caption{(Color online) The same as in Fig.~\ref{Fig_Ca48sp}
for~$^{50}$Ca.}
\label{Fig_Ca50sp}
\end{figure}

In order to construct the wave functions~(\ref{wf2ph}) of the
$1^{-}$~states, in the present study we take into account all
two-phonon terms built from the phonons with
multipolarities~$\lambda{\leq}5$~\cite{Arsenyev12,Arsenyev15,Arsenyev17}.
As an example the energies and reduced transition probabilities of
the first~$2^{+}$ and $3^{-}$~phonons for
$^{48,50}$Ca are presented in Figs.~\ref{Fig_Ca48sp} and \ref{Fig_Ca50sp}. The QRPA
results obtained with the SLy5 EDF are compared with the
experimental data~\cite{Burrows06,Elekes11}.
The $E1$~transition matrix elements are calculated with
the effective neutron, $e_{({\rm eff})}^{({\rm n})}{=}-\frac{Z}{A}e$, and proton,
$e_{({\rm eff})}^{({\rm p})}{=}\frac{N}{A}e$, charges.
Inclusion of the effective charges eliminates
contaminations of the physical response due
to the spurious excitation of the center of mass.
In Ref.~\cite{Arsenyev10}, it has been shown
that eliminating the spurious state by means
of effective charges or the alternative ways
(see, e.g., Ref.~\cite{Colo00})
leads to very close results.
As one can see, the overall agreement of the energies and
$B(E\lambda;\downarrow)$ values with the data looks reasonable.
We should be noted that calculations with the forces SLy5
are in good agreement with the values calculated with the DF3a EDF
in Ref.~\cite{Saperstein16}.

Calculating the electric dipole strength function we include
in the model wave function~(\ref{wf2ph}) all one-phonon dipole states
with energies below 35~MeV and 15~most collective phonons of
other multipolarities. The effect of configuration space extension on
the results was tested and its minor role was found.

\section{Results and discussion}

\begin{figure}[t!]
\includegraphics[width=\columnwidth]{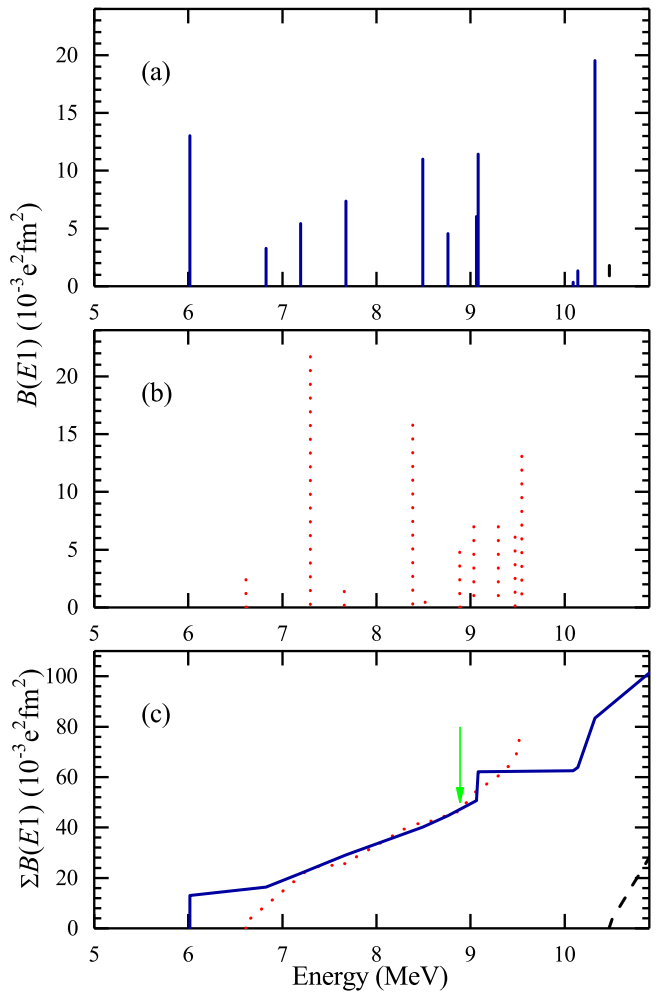}
\caption{(Color online) $1^{-}$~states  at $E_{x}{<}11$~MeV in~$^{48}$Ca.
Panel~(a) --- $1^{-}$~states calculated within RPA (a small stick
in the right corner of the panel) and ``RPA{+}PPC'' (solid sticks).
Panel~(b) --- experimental data on electric dipole states
from~\cite{Hartmann04,Derya14}. Panel~(c) ---
calculated (dashed and solid lines) and experimental (dotted line)
running sums $\sum{B(E1)}$ of electric dipole strength at~$E_{x}{<}11$~MeV.
Calculations are performed with the Skyrme force SLy5 within RPA and
``RPA{+}PPC'' aprroaches. The arrow indicates the calculated~$S_{\rm 1n}$ energy.}
\label{Fig_Ca48BE1}
\end{figure}
As the first step, we examine the PPC
effect on the $E1$~strength distributions in~$^{48,50}$Ca isotopes.
It is noteworthy that for all the $E1$~distributions,
considered in the present paper, the matrix elements
of direct transitions from
the ground state (the phonon vacuum) to two-phonon components
are about two orders of magnitude
smaller relative to ones for the excitation of
one-phonon components~\cite{Voronov84}. Thus, the transitions
from the ground state into the two-phonon part
of wave functions~(\ref{wf2ph}) are omitted in our analysis.
Inspired by Refs.~\cite{Arsenyev17,Egorova16}
we consider as the PDR in~$^{48,50}$Ca the dipole states placed below 10~MeV.
Let us now discuss the total value of the $E1$~strength concentrated  in
this energy interval. A comparison with
recent experimental data~\cite{Hartmann04,Derya14} for~$^{48}$Ca
shows that the RPA approach cannot reproduce the low-energy part of
the $E1$~strength distribution (see Fig.~\ref{Fig_Ca48BE1}).
According to RPA calculations the lowest dipole state has the energy around 10.5~MeV. In
contrast to the RPA calculations, the inclusion of the
two-phonon terms results in the formation of several
$1^{-}$~states in this energy region (see Fig.~\ref{Fig_Ca48BE1}).
The dominant contribution to their
wave functions comes from the two-phonon
configurations (${>}60{\%}$). Their one-phonon parts originate from the
fragmentation of the RPA states lying above 10~MeV.
As one can see in Fig.~\ref{Fig_Ca48BE1}(c), the calculated value of the running sum
$\sum{B(E1)}$ is close to the experimental
value. The PPC calculations give a total value of the dipole strength equals to
0.063~e$^2$fm$^2$ (the summation
includes all the dipole states below 10~MeV). The experimental value
of~$\sum{B(E1)}$ is
$0.0687(75)$~e$^2$fm$^2$ according to \cite{Hartmann04} and
$0.080(8)$~e$^2$fm$^2$ according to~\cite{Derya14} for the same
interval. Thus, the PPC effects produce a strong
impact on the low-energy $E1$ strength in~$^{48}$Ca.
As concerns other theoretical results, the calculations within
the relativistic quasiparticle time blocking
approximation~(RQTBA) estimate the value of
$\sum{B(E1)}$ as 0.1~e$^2$fm$^2$~\cite{Egorova16}.

The photon scattering experiments $^{48}$Ca$(\gamma,\gamma^{\prime})$
allow to determine the sum of energy-weighted $E1$ strength.
According to~\cite{Hartmann04} $0.33(4)${\%} of the TRK sum rule
can be attributed to the PDR region of $^{48}$Ca.
The corresponding RQTBA result is 0.55{\%}~\cite{Egorova16},
whereas our calculations with PPC effects give 0.28{\%}.

\begin{figure}[t!]
\includegraphics[width=\columnwidth]{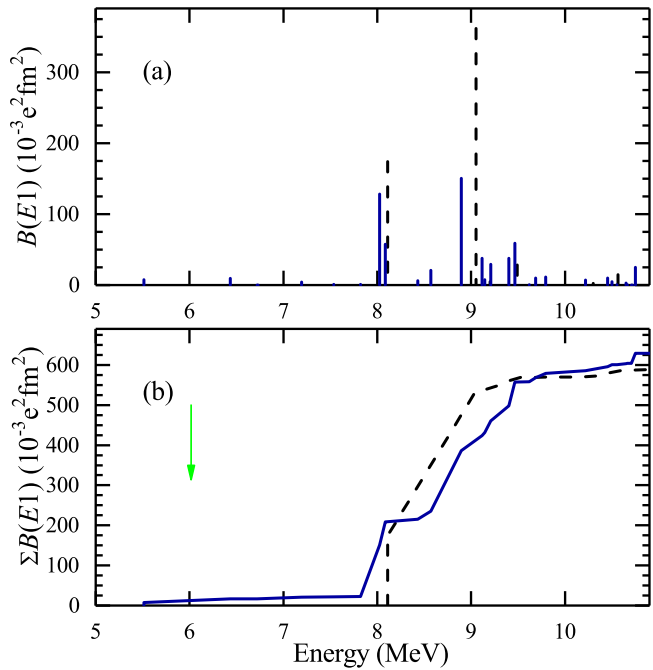}
\caption{(Color online) Theoretical results for $1^{-}$~states
at $E_{x}{<}11$~MeV in~$^{50}$Ca. Panel~(a) ---
energies and $B(E1)$-strengths of $1^{-}$~states calculated with
the Skyrme force SLy5 within QRPA (dashed sticks) and ``RPA{+}PPC''
(solid sticks). Panel~(b) --- running sums of the dipole strength
$\sum{B(E1)}$~calculated with the Skyrme force SLy5
within QRPA (dashed line) and ``RPA{+}PPC'' (solid line).
The arrow indicates the calculated $S_{\rm 1n}$.}
\label{Fig_Ca50BE1}
\end{figure}

Moving from~$^{48}$Ca to~$^{50}$Ca, the QRPA calculations predict
a jump of the $\sum{B(E1)}$ value. The neutron number $N{=}30$ corresponds to the
occupation of the neutron $2p\frac{3}{2}$ subshell, resulting in appearing the two
rather pronounced $1^-$ states below 10~MeV. These two states are practically pure neutron
2QP excitations $99{\%}\{3s\frac{1}{2}2p\frac{3}{2}\}_{\rm n}$
and $98{\%}\{2d\frac{5}{2}2p\frac{3}{2}\}_{\rm n}$.
The contribution from the~2QP proton components is invisible.
As can be seen from Fig.~\ref{Fig_Ca50BE1}, it is these states determine
the value of $\sum{B(E1)}=$0.54~e$^2$fm$^2$.

Thus, the structure of the lowest one-phonon states in $^{50}$Ca is very different
from that of the lowest $1^{-}$ RPA state in $^{48}$Ca. The latter is mainly
a proton  state built from 2QP configuration
$\{2p\frac{3}{2}1d\frac{3}{2}\}_{\rm p}$ giving a contribution of
96{\%}. In~$^{48}$Ca, the closure of the neutron subshell
$1f\frac{7}{2}$ leads to vanishing of the neutron pairing and
the $B(E1;0^{+}_{gs}{\rightarrow}1^{-}_{1})$ value is exhausted by the proton
2QP configurations. The main difference between
the two isotopes is that the neutron 2QP configurations contribute
more than proton ones in~$^{50}$Ca. This circumstance is mainly responsible for
the~$\sum{B(E1)}$ increase (e.g., Ref.~\cite{Arsenyev17}).
Calculations with the forces SLy5 and SLy5{+}T do not change
the above conclusion.

The PPC only slightly affect the summed $E1$ strength below 10~MeV in~$^{50}$Ca
(see the panel (b) in Fig.~\ref{Fig_Ca50BE1}).
The phonon-phonon coupling  mainly produces the fragmentation of $E1$~strength
among different states, Eq.~(\ref{wf2ph}), and the energy shift of dipole spectrum
in the low-energy part. The main contribution in the wave function structure
of the first $1^{-}$~state in  $^{50}$Ca comes from the two-phonon configuration
$[2_{1}^{+}{\otimes}3_{1}^{-}]_{QRPA}$ ($76{\%}$).
This configuration dominates in the structure of
the first $1^{-}$~state in~$^{48}$Ca as well.

We found that one-phonon components dominate
in the structure of four dipole states in~$^{50}$Ca ---
$1_{7}^{-}$, $1_{11}^{-}$, $1_{13}^{-}$ and $1_{16}^{-}$.
The contribution of these components in the norms of the aforementioned
states is greater than~86{\%}. These four states give the main contribution to
the summed $E1$~strength below 10~MeV.
The dominant contribution to the wave
functions of other $1^{-}$~states
comes from the two-phonon configurations (${>}75{\%}$).
We got a total dipole strength of 0.57~e$^2$fm$^2$ for the QRPA and
0.58~e$^2$fm$^2$ for the PPC calculations. The
RQTBA result is somewhat less --- 0.46~e$^2$fm$^2$~\cite{Egorova16}).

\section{Summary}

Starting from the Skyrme mean-field calculations and QRPA,
the distributions of the electric dipole strength in $^{48,50}$Ca
were studied by taking into account the coupling
between one- and two-phonons terms in the wave functions of
excited states. The finite-rank separable approach for the QRPA
calculations enables one to reduce remarkably the dimensions of
the matrices that must be inverted to perform nuclear structure
calculations in very large configuration spaces.

Neutron excess effects on the PDR excitation energies and
transition strengths were investigated.
The impact of the shell closure $N{=}28$ on
the summed $E1$~strength below 10~MeV was found.
The summed dipole transition strength $\sum{B(E1)}$ in
the PDR region noticeably increases after the crossing
the neutron shell $N{=}28$. At the same time
the one-neutron separation energy decreases by~64{\%}
in~$^{50}$Ca in compare with~$^{48}$Ca.
The latter is a result of the pairing effect on neutron
$2p\frac{3}{2}$ subshell for~$^{50}$Ca.

The present model can be extended by complicating
the trial model function of
dipole states by adding three-phonon configurations.
The computational developments that would allow us
to conclude on this point are underway.

\section*{Acknowledgments}

N.N.A., A.P.S., and V.V.V. thank the hospitality of
INP-Orsay where a part of this work was done.
This work was partly supported by the CNRS-RFBR agreement
No. 16-52-150003, the IN2P3-JINR agreement,
and the RFBR grant No. 16-02-00228.

\end{document}